# PVLAS : probing vacuum with polarized light

E. Zavattini[1], G. Zavattini[2], G. Ruoso[3], E. Polacco[4], E. Milotti[5], M. Karuza[1], U.Gastaldi[3*], G. Di Domenico[2], F. Della Valle[1], R. Cimino[6], S. Carusotto[4], G.Cantatore[1] and M. Bregant[1]

1-INFN Sezione di Trieste and University of Trieste,Italy, 2-INFN Sezione di Ferrara and University of Ferrara, Italy, 3-INFN Laboratori Nazionali di Legnaro, Legnaro (Pd), Italy, 4-INFN Sezione di Pisa and University of Pisa, Italy, 5-INFN Sezione di Trieste and University of Udine, Italy, 6-INFN Laboratori Nazionali di Frascati, Frascati, Italy

**Abstract**

The PVLAS experiment operates an ellipsometer which embraces a superconducting dipole magnet and can measure ellipticity and rotation induced by the magnetic field onto linearly polarized laser light. The sensitivity of the instrument is about $10^{-7}$ rad Hz$^{-1/2}$. With a residual pressure less than $10^{-7}$ mbar the apparatus gives both ellipticity and rotation signals at the 10-7rad level with more than 8 sigma sob ratio in runs that last about 1000 sec. These signals can be interpreted as being generated largely by vacuum ellipticity and dichroism induced by the transverse magnetic field. If this interpretation is correct, a tool has become available to characterize physical properties of vacuum as if it were an ordinary transparent medium. A microscopic effect responsible for this induced dichroism could be the existence of ultralight spin zero bosons with mass of the order of $10^{-3}$ eV, that would couple to two photons and would be created in the experiment by interactions of photons of the laser beam with virtual photons of the magnetic field. The inverse of the coupling constant to two photons would correspond to a mass M of the order of $10^6$ GeV.

Keywords: axion; boson; dark matter; dichroism; birefringence; polarization

## 1. Introduction

The polarization of light propagating in vacuum through a transverse magnetic field can be modified by interactions of photons of the light beam with virtual photons of the magnetic field.

Photon-photon interactions producing a real particle deplete the original light beam. If e.g. a pseudoscalar particle is produced only photons with the electric field direction parallel to the magnetic field are absorbed. The component of the polarized light beam with polarization parallel to the magnetic field is reduced, while the orthogonal component remains unchanged and then the outcoming light beam has the polarization plane slightly rotated (vacuum magnetic induced dichroism). The rotation angle is proportional to $\sin 2\theta$, ($\theta$ is the angle between the magnetic field and the polarization vector).

Photon-photon interactions mediated by virtual particles or fermion loops do not change the number of photons of the light beam, but may introduce different phase delays for the two polarization components. This generates a small ellipticity (vacuum magnetic induced birefringence), that is proportional to $\sin 2\theta$.

Observation and measurements of amplitude and phase of ellipticity and dichroism induced in vacuum by a trasverse magnetic field can be used to characterize quantum vacuum and to discover the existence of ultralight particles with mass below the energy of the photons of the light beam, if enough sensitivity can by obtained from the experimental apparatus.

In 1979 a method was suggested to observe QED photon–photon interactions from the measurement of vacuum birefringence [1], and in 1986 it was shown that mass, coupling and parity of an ultralight spin-zero particle could be determined from measurements of both vacuum dichroism and birefringence [2]. At BNL a pilot experiment of this kind has given limits on the coupling to two gammas of light spin-zero bosons [3].

Several years of developments [4-21] have led to the present configuration of PVLAS. The experiment operates at LNL an ellipsometer with a sensitivity of some $10^{-7}$ rad Hz$^{-1/2}$ that is capable of measuring, with the use of the heterodyne technique, amplitude and phase of both ellipticity and dichroism. It has started data taking in 2001 and since some years observes clear peaks at the frequencies of the Fourier spectrum of the detector response where authentic dichroism and ellipticity signals should appear. The amplitude of these peaks (measured with residual pressure less than $10^{-7}$ mbar and 1064nm wavelength laser light until all 2004) is at the $10^{-7}$ rad level and is well above the noise in the side bins of the frequency spectrum (the signal over background ratio is usually better than 8 for runs lasting more than 600 sec). The amplitude of these peaks is rather surprising, because (a) the candidate ellipticity signals exceed by 4 orders of magnitude expectations for QED induced ellipticity and (b) the candidate dichroism signals imply a coupling constant 1/M of about $10^{-6}$ GeV$^{-1}$ for an axion-like particle of mass m around $10^{-3}$ eV, that is four orders of magnitude larger than recent limits obtained by CAST [22].

A long sequence of tests and measurements has then been performed to establish the physical sources of the observed peaks. The analysis of dichroism data has led so far to the conclusion that both a vacuum dichroism effect and a background apparatus effect are present [19-21].

Under the hypothesis that a large component of the signals in the relevant peaks of the Fourier spectrum are genuine vacuum effects, and that the microscopic source of the dichroism effect is the existence of a very light boson, the excursion of the values of the dichroism and ellipticity signals define two bands in the (m,M) plane that intersect in an area surrounding the point at m=$10^{-3}$eV and M=$10^6$ GeV.

This conclusion is strongly supported by the results of the analysis of data collected with the addiction of small quantities of noble gas at various pressures in the apparatus [19]. Photon low-mass boson mixing is expected to feature a diffraction-like behaviour [23] that can be modified in a predictable way by reducing the speed of photons by increasing the refraction index of their propagation medium [23,24]. Data seem to really follow the predicted pressure dependence. Furthermore we have recently realized that the background effect that adds vectorially to the vacuum dichroism signal, has a nearly constant effect on the total signal amplitude in dichroism measurements with gas at relatively high pressures. This point strengthens the validity of the analysis of ref [19], that exploited the pressure dependence of the dichroism peak amplitude in the pressure window 1-20 mbar of Ne.

If confirmed, the PVLAS signals are obviously relevant for axion and dark matter studies (for recent reviews and references see refs. [25-27]), but might as well be originated by other new physics (see e.g. ref [28]).

In May 2005 we have started collecting data with 532nm green light from a frequency doubled laser. Ellipticity and dichroism signals are present also in these new data. The scenario with green light seems to reproduce the scenario with infrared light. A quantitative analysis of the data is in progress.

For the near future it is foreseen the installation of an access structure in aluminum, that will reduce mechanical and magnetic couplings between the magnet and the optics.

Modular permanent magnets would permit to vary the length of the magnetized region and give the bonus of a much better duty cycle and of no stray field. Use of permanent magnets is envisaged in order to confirm the nature of the ellipticity and rotation signals and to improve the accuracy the determination of m and M.

Among the various suggestions to exploit interactions of ultralight or zero mass bosons with electromagnetic fields in order to ascertain their existence [2,29-33], a regeneration experiment [30-32] of the type performed in ref. [34] would be extremely convincing, if it would give a positive answer. A precise determination of the m and M parameters with PVLAS would permit to optimize the design of such a regeneration experiment. A regeneration extention positioned above or below PVLAS could also be imagined if major logistics difficulties could be overcome.

In the following sections we cover briefly apparatus, experimental method, calibrations and ellipticity and dichroism measurements.

## 2. Apparatus and experimental method

A schematic drawing of the PVLAS set up is given in fig.1 of refs.[18]. Polarized light propagates in vacuum along a vertical axis and traverses a magnetic field of length L=1m generated by a superconducting dipole magnet that is operated at a field B up to 5.5 T. The magnetic field lines are horizontal. The magnet can rotate around a vertical axis that coincides with that of light propagation. In order to increase the optical path through the magnet, the light reflects between two mirrors M1 and M2 installed respectively below and above the magnet. The distance d between M1 and M2 is d=6.4 m. The two mirrors form a Fabry-Perot optical cavity. Light undergoes N reflections on average, and N is of the order of $10^5$. Linearly polarized light is injected into the FP cavity through a polarizer P1 located below the magnet. The orientation of P1 is kept fixed for all the measurements. The intensity of the induced ellipticity and dichroism is proportional to $\sin 2\theta$. The ellipticity and dichroism effects have maximum amplitude when the magnetic field is oriented at $45^0, 135^0, 225^0$ and $315^0$ to the polarizer, and are null for θ values $0^0, 90^0, 180^0$ and $270^0$.

A polarizer P2 identical to P1 is mounted above the magnet behind M2 and is oriented orthogonal to P1 with a fine adjustment in order to minimize the transmission with magnet not powered of the light coming from the cavity (maximum extinction). A photodiode D detects the light emerging from P2. Ideally it should not give signals when the magnet is off and when it is on, but it is oriented with θ at $0^o$, $90^o$, $180^o$ and $270^o$, while it should give signals of maximum amplitude when the magnet is on and it is oriented with θ at $45^o$, $135^o$, $225^o$ and $315^o$.

In order to enhance the signal over background ratio a heterodyne technique is used. For this purpose a small ellipticity of amplitude about $10^{-3}$rad is given to the light beam that emerges from M2 by an electrooptical modulator SOM [14] that is driven by a precise sine pulse generator at a frequency $\omega_s$=506 Hz. With the magnet set in rotation at a frequency $\omega_m$, the ellipticity and dichroism effects generated when the light passes through the magnet bore are modulated at $2\omega_m$. The angular position of the magnet is measured by a set of 32 marks on the periphery of the supporting turntable. The identifier of each mark and its time of passage through a measuring station are also continuously recorded, so that nonuniformities of the rotation speed of the magnet can be corrected accurately in Fourier analysis of the data [17].

For dichroism measurements a quarter wave plate QWP is inserted between M2 and SOM with the QWP principal axis aligned with the SOM main axis. The QWP transforms rotations originated in the FP cavity into ellipticities and viceversa. The QWP axis can be rotated by $90^o$. This operation causes a $180^o$ phase rotation of an optical signal [35]. The data acquisition system collects with a frequency of 8.2 KHz signals from the photodiode D (after amplification and processing by a 24 bit ADC), from the system that measures the magnet orientation and from normalization and control devices.

The output of the photodiode is Fourier analyzed to search for ellipticity (or dichroism transformed into an ellipticity by the QWP) signals. The tiny effect of the ellipticity (or dichroism) induced by the magnetic field and modulated at twice the rotation frequency of the magnet $2\omega_m$ beats with the small (but much larger) ellipticity introduced by the SOM at $\omega_s$ and generates signals in the two $\omega_s \pm 2\omega_m$ sidebands of $\omega_s$[10]. We know therefore the frequencies where the signal is expected and also its phase (apart an incertitude of 180°), because the magnet effect is maximal both at 45° and at 225° from the polarizer. A very practical and convenient way to calibrate the amplitude and monitor the direction where the signal should appear in a polar plot (where the amplitude is given by the radius and the phase by the angular direction) without tracking all the offsets in the apparatus and in the software, consists in injecting gas in the optical path. Ellipticity is induced by the magnetic field in the magnetized volume of the gas by Cotton-Mouton effect (see refs.[16,18] and refs. therein) with an amplitude that is proportional to the gas pressure and phase that is independent on pressure and gives the required angle.

## 3. Ellipticity and dichroism measurements

We have performed several measurements under different experimental conditions in order to understand the sources of the signals in the $2\omega_m$ side bands of the $\omega_S$ peak of the Fourier spectrum of the output of the signal photodiode. We have observed or verified experimentally the following facts.

The $2\omega_m$ peak amplitude is independent on the frequency of rotation of the magnet in ellipticity measurements with gas in the FP.

The $2\omega_m$ peak amplitude scales as the finesse of the Fabry-Perot resonator.

With the magnet not powered, but set in rotation, the $2\omega_m$ signal is no more present.

With the mirrors M1 and M2 of the FP cavity removed the $2\omega_m$ signal disappears.

With the bottom mirror M1 moved upwards near to the mirror M2 to make a shorter FP cavity above the superconducting magnet, no $2\omega_m$ signal has been detected.

The $2\omega_m$ amplitude of the ellipticity signal generated by Cotton–Mouton effect by introducing gas in the magnetized optical path is proportional to the gas pressure (see fig. 3 in ref.[18]).

The phase of the $2\omega_m$ signal with gas in the magnetized optical path is independent on the gas pressure (see fig. 6 in ref.[19]), and corresponds, as expected, to an angle of 45° between the magnetic field direction and the axis of the polarizer.

The phase of the $2\omega_m$ ellipticity signal follows a rotation of the polarizer.

At a fixed gas pressure the amplitude of the ellipticity signal with gas in the magnetized optical path is proportional to $B^2$.

With vacuum in the optical cavity, by varying the magnetic field intensity B, the $2\omega_m$ dichroism signal amplitude is compatible with a $B^2$ dependence, and the phase remains stable for measurements performed sequentially the same day without accessing the optics.

Pairs of sequential runs in vacuum with QWP inserted taken respectively with the QWP 0° and QWP 90° settings feature $2\omega_m$ signals with comparable amplitudes and phases differing by about 180° (see fig.3 in ref.[20]) in agreement with expectations for the behaviour of optical rotation signals [35].

Different periods of data taking in vacuum feature generally different amplitudes and phases. Dichroism amplitudes vary from $10^{-7}$ rad to $4 \cdot 10^{-7}$ rad with B=5.5 T and N=50.000. Instead, by splitting runs that last longer than 600 sec into e.g. 100 sec long samples, the amplitudes and phases of the samples cluster regularly around a central value with a small spread (see fig.3 in ref.[20]). The ratio between the signal amplitude of $2\omega_m$ peaks and the background signal in side bins is typically better than 5 for runs lasting more than 6oo sec (see fig 7 in ref.[19] and fig. 2 in ref. [21]). The statistical significance of the $2\omega_m$ peaks is therefore not a problem. The main issues are instead the phases of the $2\omega_m$ signals and what are the sources of the $2\omega_m$ signals. It is very encouraging that nearly all of the 0° QWP runs have their phase in one half of the phase plane and nearly all of the 90° QWP runs have their phases in the complementary half (see fig 3 in ref. [21]). This is consistent with the presence of a vacuum signal with phase aligned along the axis indicated by the calibration ellipticity signals plus a background signal with phase and amplitude which vary from period to period (possibly as the result of two components, one on the optical axis and the other orthogonal to it, that have relative amplitudes subject to variations from period to period).

The ensemble of the results mentioned above is compatible with the existence of vacuum induced ellipticity and dichroism together with instrumental effects which add vectorially to generate the $2\omega_m$ sidebands of $\omega_s$.

Nethertheless the results mentioned above are not sufficient to exclude that a systematic effect of the apparatus with phase on the optical (ellipticity) axis could mimic vacuum effects. This possibility seems however excluded in the case of dichroism. Indeed the hypothesis that vacuum dichroism is present and observable in our apparatus is strongly supported by the results of a detailed scrutiny of the pressure dependence of the amplitude of the peaks at $\omega_s \pm 2\omega_m$ with QWP inserted measured as a function of increasing gas pressure in a sequence of runs following a run with vacuum.

Fig 1 shows row data of ellipticity and dichroism measurements performed during october 2004 in vacuum and at various increasing pressures of Ne from 1 to 21 mbar . Gas measurements were executed in strings of sequential runs by increasing the pressure in the optical cavity between data taking runs and no other intervention on the apparatus.

By adding gas in sequential runs no actions are made on the optics and on the apparatus and background effects are expected to remain stable. The gas generates ellipticity by CM effect, that is filtered out by the QWP plate in dichroism measurements. However a fraction of the ellipticity signal, which is proportional to the gas pressure, leaks through the QWP due to non ideal optical elements and becomes a component of the dichroism signal. This effect is clearly noticeable in fig 1.

The addition of gas in the FP cavity, besides generating the CM effect, increases the index of refraction of the medium where light propagates from 1 to n, and slows then down the light propagation in the FP cavity, while it does not affect the propagation of ultralight spin zero bosons, that have quasi null coupling to matter. The dependence of light-boson photon mixing in vacuum on the mass m is described in the (m,M) plane by a diffraction-like curve (see fig.9 of ref. 19). When gas is introduced in the light path, the effect of the pressure dependence of the refraction index n is equivalent to a pressure dependent change of the light-boson mass m. By changing the pressure p one runs therefore along the diffraction curve in the (m,M) plane. If the noise level is low enough, with increasing pressure one expects then to observe, after subtraction of a constant background amplitude and of the pressure dependent CM feedthrough, a decrease of the dichroism signal down to zero followed by a damped oscillating behaviour,. This is indeed visible in the plot of the candidate dichroism signal shown in fig. 10 of ref. [19] in function of the pressure of Ne gas. A fit of the experimental curve with m and M as free parameters gives values inside the region of the (m,M) plane identified by the crossing of the ellipticity and dichroism bands.

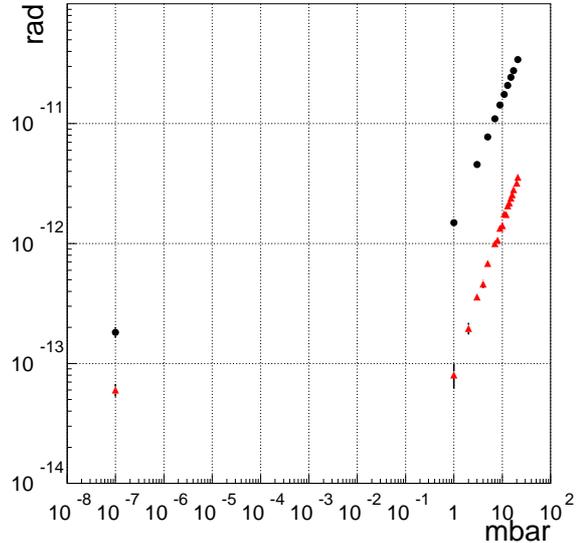

*Fig.1 Ellipticity (round points) and dichroism (triangles) measured with vacuum (<$10^{-7}$ mbar) and Ne gas (pressures are in mbar) in the optical path. The ordinate (in radians) gives the effect per transit of the polarized light through one meter of 1T transverse magnetic field.*

At zero pressure the vacuum dicroism signal is maximum, but it cannot be observed directly because its signal adds vectorially with the total background effect of the apparatus, and it is not sufficient to project the signal vector along the optical axis in the phase plane in order to remove the background signal, since the background effect may have a component along the optical axis.

The dichroism signal component generated by ellipticity feedthrough is aligned with the optical axis, as well as the genuine vacuum dichroism component, which must be aligned, if it exists, along the optiacal axis not only in vacuum, but also at all pressures.

At low pressures the phase of the dichroism signal should move rapidly if the phase and amplitude of the background term (BT) stay constant, because the phases of the vacuum dichroism (VD) and gas CM feedthrough (CMft) stay on the optical axis, but the amplitude of the combined VD+CMft signals changes rapidly, due to the pressure change of the CMft term.

At intermediate and (relatively) high pressures the phase of the total dichroism signal tends to align with the optical axis and the amplitude of the signal is dominated by the amplitude of the total component CMft+VD on the optical axis. Therefore the vacuum contributions (that are affected by the index of refraction effect) obtained in ref. [19] by subtraction of the linear CMft term from the total amplitude were marginally affected by neglecting the background term BT.

If the microscopic origin of ellipticity and dichroism is dominantly due to the existence of a light boson of mass m which couples to two photons with an inverse coupling constant M, the crossing in the (m,M) plane of the two bands of equiellipticity and equidichroism curves identifies a relatively small oasis in the (m,M) plane around $m=10^{-3}$eV and $M=10^{+6}$GeV for the mass and coupling parameters of the light boson. This oasis is deeply inside the desert of the exclusion plot mapped by axion search experiments [22,25-27] in the (m,M) plane.